\documentclass[twocolumn,aps,prc,superscriptaddress,reprint,floatfix]{revtex4-1}
\usepackage{graphicx}
\usepackage[latin2]{inputenc}
\usepackage[T1]{fontenc}
\usepackage{wrapfig}
\usepackage{titlesec}
\usepackage{amsmath}
\usepackage{bm}% bold math
\setcitestyle{numbers,square}
\usepackage{hyperref}
 \hypersetup{
     colorlinks=true,
     linkcolor=black,
     filecolor=blue,
     citecolor = blue,      
     urlcolor=cyan,
     }

\begin{document}

\titlespacing\section{0pt}{20pt plus 4pt minus 2pt}{3pt plus 20pt minus 2pt}
\titleformat{\section}
  {\normalfont\fontsize{10}{10}\bfseries\centering}{\thesection}{0.5em}{}
\renewcommand\thesection{\Roman{section}.} 
%*** ADD YOUR TITLE HERE ***%

\title{\bf Fusion of $^{16}$O+$^{165}$Ho at deep sub-barrier energies}

\affiliation{
Saha Institute of Nuclear Physics, 1/AF, Bidhan Nagar, Kolkata - 700064, 
India}
\affiliation{Nuclear Physics Division, Bhabha Atomic Research Centre, Mumbai - 400085, 
India}
\affiliation{Homi Bhabha National Institute, Anushaktinagar, Mumbai - 400094, India}
\affiliation{Tata Institute of Fundamental Research, Mumbai - 400005, India}
\affiliation{Vivekanand Education Society's College of Arts, Science \& Commerce,
Mumbai - 400071, India }

\author{Saikat Bhattacharjee}
\affiliation{
Saha Institute of Nuclear Physics, 1/AF, Bidhan Nagar, Kolkata - 700064, 
India}
\affiliation{Homi Bhabha National Institute, Anushaktinagar, Mumbai - 400094, India}
\author{A. Mukherjee} 
\altaffiliation {anjali.mukherjee@saha.ac.in} %FOR CORRESPONDING AUTHOR
\affiliation{
Saha Institute of Nuclear Physics, 1/AF, Bidhan Nagar, Kolkata - 700064, 
India}
\affiliation{Homi Bhabha National Institute, Anushaktinagar, Mumbai - 400094, India}
\author{Ashish Gupta}
\affiliation{
Saha Institute of Nuclear Physics, 1/AF, Bidhan Nagar, Kolkata - 700064, 
India}
\affiliation{Homi Bhabha National Institute, Anushaktinagar, Mumbai - 400094, India}
\author{Rajkumar Santra}
\affiliation{
Saha Institute of Nuclear Physics, 1/AF, Bidhan Nagar, Kolkata - 700064, 
India}
\affiliation{Homi Bhabha National Institute, Anushaktinagar, Mumbai - 400094, India}
\author{D. Chattopadhyay}
\altaffiliation{Present adddress: Tata Institute of 
Fundamental Research, Mumbai - 400005, India}
\affiliation{
Saha Institute of Nuclear Physics, 1/AF, Bidhan Nagar, Kolkata - 700064, 
India}
\author{N. Deshmukh}
\altaffiliation{Present address: School of Sciences, Auro University, Surat, 
Gujarat - 394510, India}
\affiliation{
Saha Institute of Nuclear Physics, 1/AF, Bidhan Nagar, Kolkata - 700064, 
India}

\author{Sangeeta Dhuri} 
\author{Shilpi Gupta}
\author{V.V. Parkar}
\affiliation{Nuclear Physics Division, Bhabha Atomic Research Centre, Mumbai - 400085, 
India}
\affiliation{Homi Bhabha National Institute, Anushaktinagar, Mumbai - 400094, India}
\author{S.K. Pandit}
\author{K. Ramachandran}
\affiliation{Nuclear Physics Division, Bhabha Atomic Research Centre, Mumbai - 400085, 
India}
\author{K. Mahata} 
\author{A.Shrivastava}
\affiliation{Nuclear Physics Division, Bhabha Atomic Research Centre, Mumbai - 400085, 
India}
\affiliation{Homi Bhabha National Institute, Anushaktinagar, Mumbai - 400094, India}

\author{Rebecca Pachuau}
\altaffiliation{Present address: Department of Physics, Banaras
Hindu University, Varanasi 221005, India.}
\affiliation{Tata Institute of Fundamental Research, Mumbai - 400005, India}
\author{S.Rathi}
\affiliation{Vivekanand Education Society's College of Arts, Science \& Commerce,
Mumbai - 400071, India }

\begin{abstract}
Fusion cross-sections have been measured for the asymmetric system 
$^{16}$O+$^{165}$Ho at energies near and deep below the Coulomb barrier with 
an aim to investigate the occurrence of fusion hindrance for the system. Fusion 
cross sections down to $\sim$ 700 nb have been measured using the off-beam $\gamma$-ray 
technique. The fusion cross sections have been compared with the 
coupled channel calculations. Although the onset of fusion hindrance could not 
be observed experimentally, an indication of a small deviation of the experimental fusion 
cross-sections with respect to the calculated cross-sections could be
observed at the lowest energy measured. However, the energy onset of fusion 
hindrance has been obtained from the extrapolation technique and is found to be
about 2 MeV below the lowest energy of the present measurement. 
\end{abstract}

\maketitle

\section{INTRODUCTION}
Extensive studies on fusion reactions at sub-barrier energies in different mass
regions have unraveled the fundamentals of quantum mechanical tunnelling and
distribution of potential barriers between two interacting nuclei. Fusion in 
the vicinity of Coulomb barrier is a probe to discern the indispensable 
role of different intrinsic features of the interacting nuclei on the reaction 
process. Enhancement of fusion cross sections observed in heavy-ion collisions 
at sub-barrier energies has been well explained by the coupled channels   
model \cite{Bala98,Das98}. 

On extending the measurements from sub-barrier down to deep sub-barrier 
energies, for a wide range of reactions \cite{cit1,cit2,cit3,Back14, 
cit4,cit5,cit6,cit7,cit8,cit9,cit10,cit11,cit12,cit13,cit14,cit15,cit16}, a 
steep fall-off of fusion excitation function has been observed compared to the 
standard coupled channels calculations, although fusion cross sections are 
still enhanced with respect to single-barrier penetration model calculations. 
This phenomenon of change of slope in the fusion excitation function at deep
sub-barrier energies is termed as "fusion hindrance". Observation of fusion 
hindrance at deep sub-barrier energies, especially in light systems like $^{12}$C+$^{12}$C 
and $^{16}$O+$^{16}$O \cite{Back14}  have 
astrophysical implications, as some of the light systems transpire in the late 
evolutionary stages of massive stars.

Fusion hindrance was initially observed in the symmetric medium-heavy system 
$^{60}$Ni+$^{89}$Y \cite{cit1}, having negative Q-value. 
Subsequently, studies of  several symmetric and nearly-symmetric 
\cite{cit2,cit3,cit4,cit5,cit6,cit7,cit8,cit9,cit10} and asymmetric
\cite{cit11,cit12} systems over a wide range of mass and fusion
Q-values have also exhibited similar observation. By contrast, fusion of weakly
bound light projectiles $^{6,7}$Li with $^{198}$Pt does not manifest hindrance 
at deep sub-barrier energies \cite{cit13,cit14}. However, in fusion with 
relatively heavier projectiles $^{11}$B and $^{12}$C on the targets $^{197}$Au 
\cite{cit15} and $^{198}$Pt \cite{cit13}, respectively, hindrance has been 
observed. From these studies, it appears that for asymmetric heavy systems,
fusion hindrance becomes increasingly significant with increasing mass and 
charge of the projectiles.

Several models with  different physical foundations have been proposed to describe this 
phenomenon. Among them, the model developed by Mi\c{s}icu and Esbensen 
\cite{Misicu2006, Misicu2007} is based on sudden approximation. A soft repulsive core 
was incorporated with the density folded M3Y potential in this model, to consider the 
nuclear incompressibility in the overlapping region of the two interacting nuclei. The 
adiabatic model proposed by Ichikawa \textit{et al.} \cite{Ichikawa2009, Ichikawa2015} 
introduces an additional damping factor on the nuclear coupling potential. The damping 
factor is a function of the internuclear distance, which takes into consideration of the 
smooth change from sudden to adiabatic transition while the two nuclei are going through 
fusion deep below the barrier. More recently, Simenel \textit{et al.} have studied the 
effect of Pauli repulsion on heavy ion fusion  by implementing the  density-constrained 
frozen Hartree-Fock method \cite{Simenel2017}. Despite having different physical origins, 
these models have been quite successful in reproducing the experimental results for 
different systems at deep sub-barrier energies \cite{Back14}. 

In the light of the problem of fusion hindrance, we have recently measured 
fusion cross-sections for $^{16}$O+$^{165}$Ho system at energies near and deep 
below the Coulomb barrier. Owing to the large deformation of nuclei in the 
rare-earth region, systems involving such nuclei usually exhibit strong 
coupling effects between relative motion and internal degrees of freedom in the
sub barrier fusion mechanism. The target nucleus $^{165}$Ho has a large 
deformation parameter \cite{cit18}, whereas the projectile $^{16}$O is a 
tightly bound spherical nucleus and therefore subdues the projectile effect of 
different coupling schemes on the fusion mechanism. It has been perceived that 
stiff systems, where coupling effects are small, typically show fusion 
hindrance more readily than soft systems \cite{Back14}. The dominance of the 
influence of different direct reaction channels, like inelastic scattering, 
transfer and breakup, in reactions involving soft nuclei are believed to be 
responsible for the occurrence of fusion hindrance at much lower energies than 
stiff systems \cite{Back14}. The system $^{16}$O+$^{165}$Ho lies between stiff 
and soft systems having a negative Q-value (Q=$-$23.1 MeV) for fusion. It 
would be interesting to study fusion mechanism in the system of tightly bound 
projectile $^{16}$O on the deformed nucleus $^{165}$Ho, especially at deep 
sub-barrier energies. 

The complete fusion(CF) cross sections of $^{16}$O+$^{165}$Ho at above-barrier 
energies have been reported in the literature \cite{cit19}. Present study 
overlaps some of the energies of the reported measurement. The measurement has been 
extended towards 
below barrier to deep sub-barrier energy region. An off-beam $\gamma$-ray 
detection technique has been implemented to measure the cross sections of the 
$\beta$-active evaporation residues. Section II recounts the detailed 
experimental method that have been implemented to perform the experiment. The procedure 
of data 
analysis have been explained in Section III. The details of theoretical coupled channel 
calculations and 
comparison with experimental results have been described in Section IV. 
Section V consists of a discussion followed by a summary of the present work 
in Section VI.

\begin{figure}[b]
\centering
\includegraphics[clip, trim=0.0cm 0.2cm -2.0cm 0.1cm,width=0.55\textwidth]{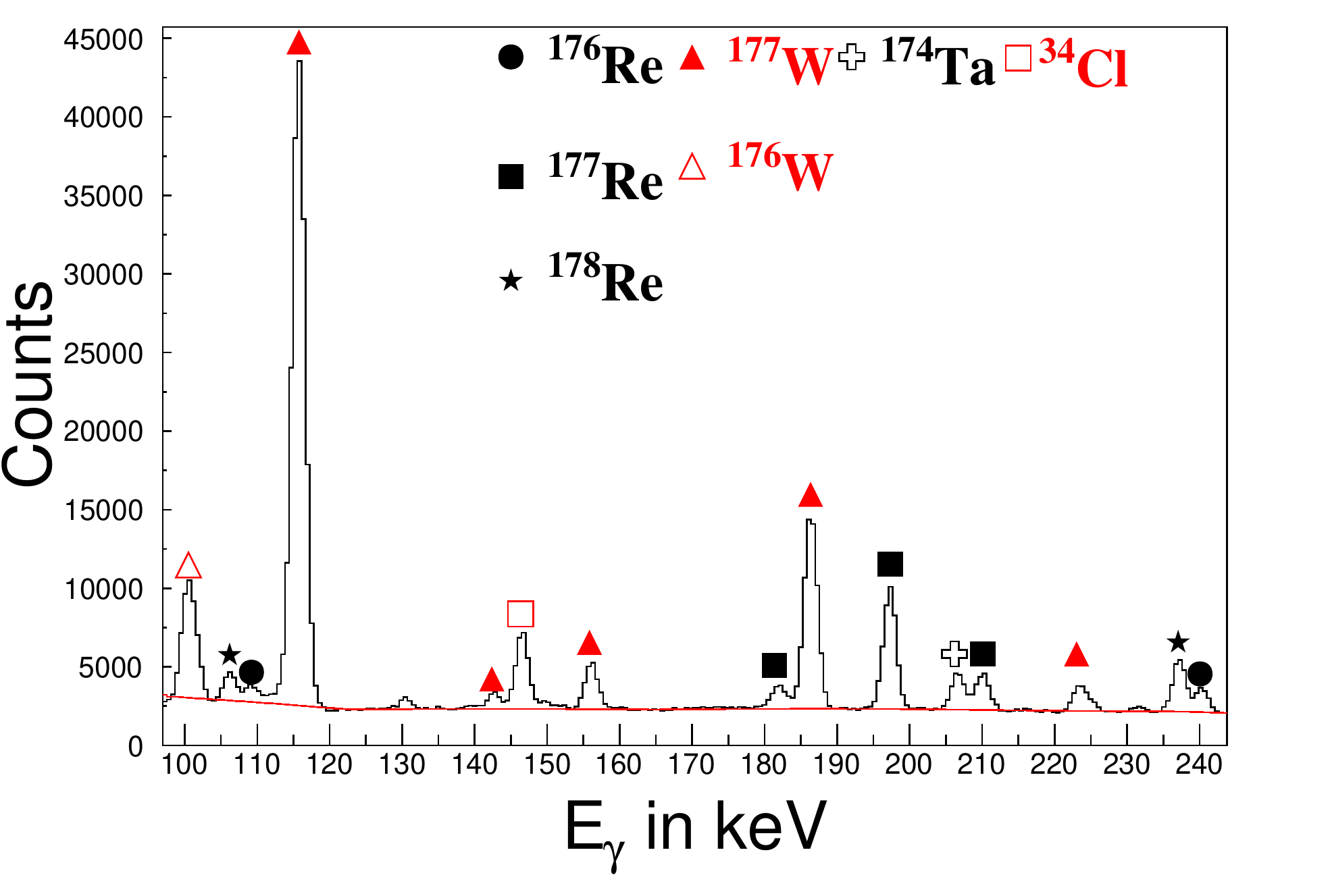}
\caption{$\gamma$-ray spectrum of the evaporation residues arising from the complete fusion(CF) 
of  $^{16}$O+$^{165}$Ho system at 84 MeV beam energy. The red line is the background estimated 
by ROOT data analysis package \cite{citnew1}. }
\end{figure}

\section{EXPERIMENTAL DETAILS}
The experiment was performed at the 14UD BARC-TIFR Pelletron-LINAC facility, Mumbai. Self-supporting, 
rolled, natural foils of $^{165}$Ho, having thickness 
in the range $\sim$ 1.02-1.9 mg/cm$^2$ were irradiated by beams of $^{16}$O, in the energy range 
E$_{lab}$=62-85 MeV. Each target foil of $^{165}$Ho was 
followed by an Al catcher foil, sufficiently thick to stop the heavy evaporation residues (ERs) 
produced in the reaction. The thickness of each target and 
catcher foil was measured by the $\alpha$-transmission method. For each irradiation, a fresh target-catcher 
foil assembly was used. Typical beam current 
during the irradiations was $\sim$ 2-10 pnA. To correct for beam fluctuations during the irradiation, the 
beam current was recorded at regular intervals of 
1 min using a  CAMAC scaler. The energies of the incident beam were corrected for the loss of energy in the 
target material by employing SRIM \cite{Ziegler2010}, at half-thickness of the target . As all the ERs were 
$\beta$-active and yielded delayed $\gamma$-rays, the activation technique was employed to 
determine the fusion cross sections for the system. After each irradiation, the target-catcher foil assembly 
was removed from the chamber and
placed in front of an efficiency calibrated HPGe detector, which detected the delayed $\gamma$-rays emitted 
by the ERs.

\begin{figure}[t]
\centering
\includegraphics[clip, trim=0.0cm 1.2cm -2.0cm 0.1cm,width=0.51\textwidth]{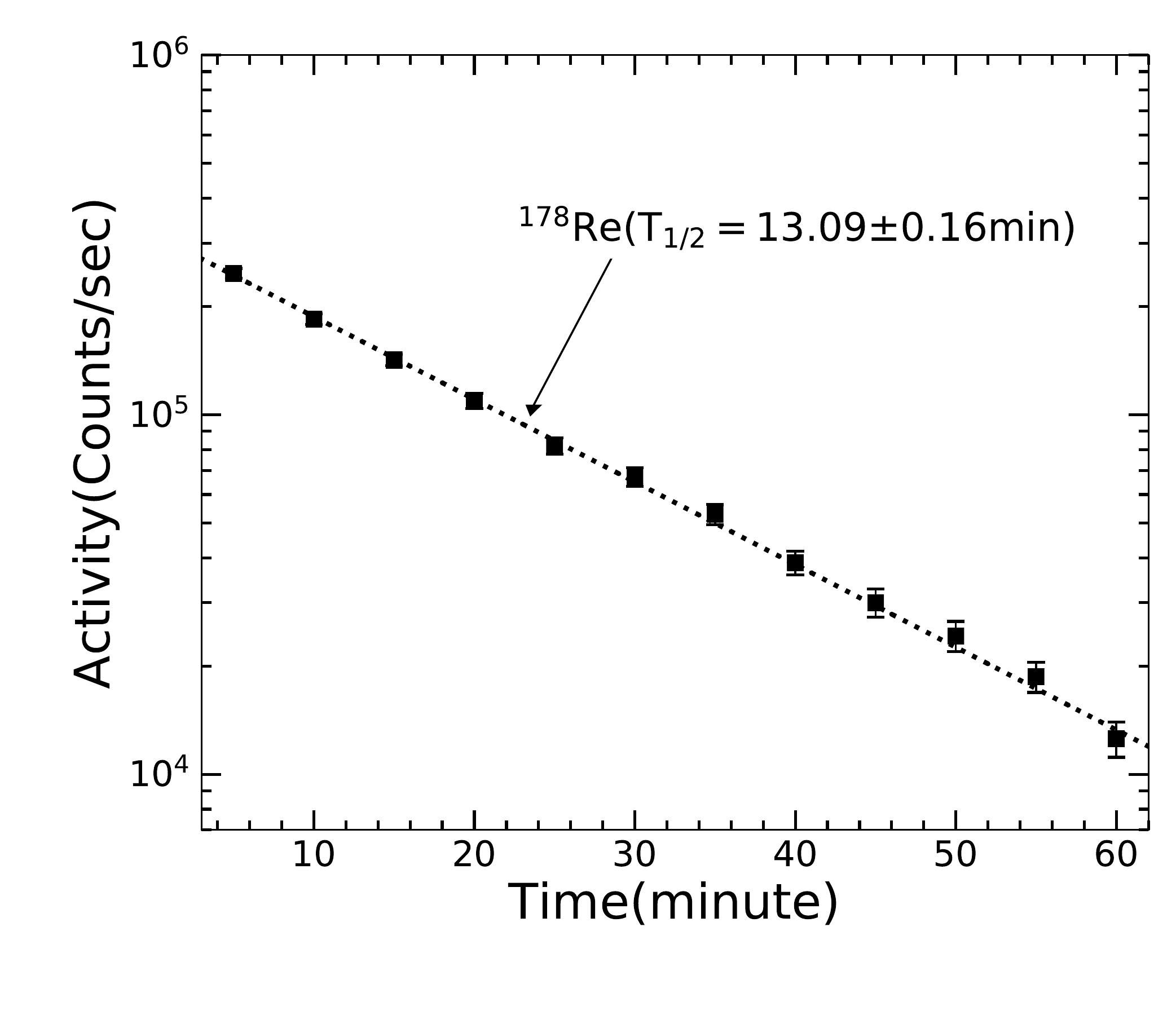}
\caption{Activity of the $^{178}$Re evaporation residue as a function of progressing time at E$_{lab}$=70 MeV. 
The dotted line corresponds to the activity obtained by fitting the data. The half life from the fit is 
mentioned in the graph, which is consistent with the known value.  }
\end{figure}

 The target-catcher foil assembly was placed either at a distance of 10 cm from the face of the detector or on the face of the detector, depending on the 
activity of the irradiated sample. The energy calibration and absolute efficiency measurement of the detector were carried out using the standard 
radioactive sources, $^{152}$Eu, $^{133}$Ba and $^{60}$Co, mounted in the same geometry as the target. The measurement was done in a low background setup 
with Pb-Cu graded shielding to reduce the background $\gamma$-rays. Data were recorded using a digital data acquisition system employing a CAEN N6724 digitizer and the 
data were analyzed using the ROOT data analysis framework \cite{citnew1}. A typical off-beam $\gamma$-ray spectrum, after the irradiation at E$_{lab}$= 84 MeV, is 
shown in Fig. 1. The complete fusion ERs $^{176-178}$Re occurring from the decay of the compound nucleus $^{181}$Re were uniquely identified from the 
characteristic $\gamma$-rays emitted by their daughter nuclei and by following the half-lives. Half lives of the ERs were measured from the time sliced 
yields of the characteristic $\gamma$-rays and compared with the previously measured values \cite{citnew2}, to ensure the absence of any contribution from 
sources other than the complete fusion residues. The half-life plot for $^{178}$Re at E$_{lab}$= 70 MeV has been shown in Fig. 2.

\section{ANALYSIS AND RESULTS}
If N$_\gamma$ represents the number of counts under a particular $\gamma$-ray peak, corresponding to a given ER in the spectrum, then from the principle of 
radioactive decay the corresponding ER cross-section ($\sigma_{ER}$) is given as \cite{cit20}:
\begin{equation}
\sigma_{ER}=\frac{N_\gamma \lambda e^{\lambda t_w}}{N_B N_T \epsilon_\gamma F_\gamma (1-e^{-\lambda t_c})(1-e^{-\lambda t_{irr}})}
\end{equation}

where N$_B$ is the number of incident nuclei, N$_T$ is the number of target nuclei per unit area, $\lambda$ is the decay constant of the ER, t$_{irr}$ is 
the irradiation time, t$_w$ is the time elapsed between the end of irradiation and the beginning of counting, t$_c$ is the counting time, $\epsilon_\gamma$ 
is the efficiency of the detector for a given $\gamma$-ray energy, and F$_\gamma$ is the absolute intensity of a $\gamma$-ray decay.    

\begin{figure}[t]
\centering
\includegraphics[clip, trim=0.5cm 0.8cm 0.3cm 0.3cm,width=0.45\textwidth]{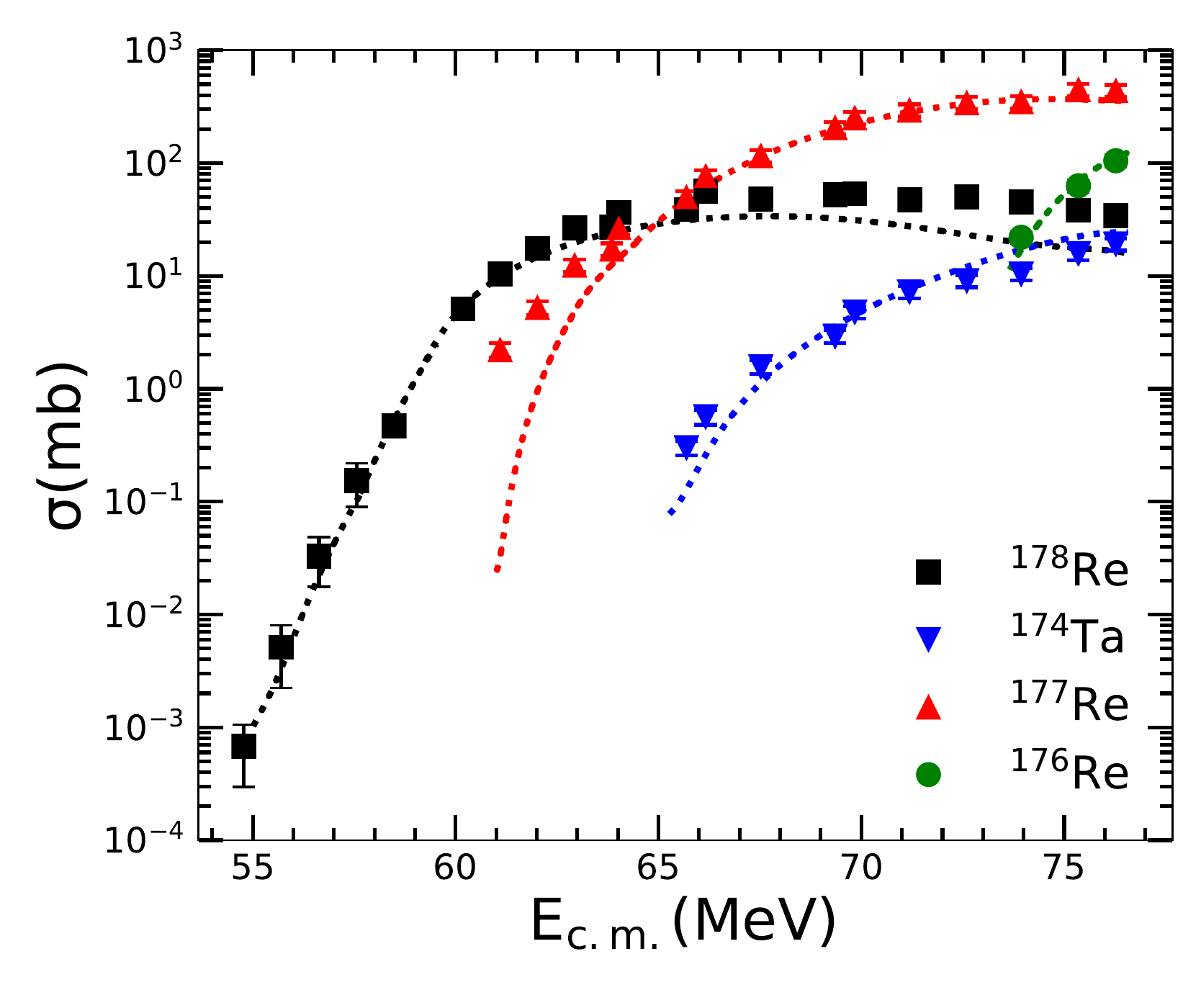}
\caption{The cross-sections of different evaporation residues of compound nucleus for $^{16}$O+$^{165}$Ho system. The dotted lines are the statistical model 
calculation (PACE4) results.}
\end{figure}
The $\gamma$-ray peak corresponding to $^{34}$Cl, seen in the spectrum (Fig. 1), arises from the reaction of $^{16}$O with the Al catcher foil. Different 
$pxn$ and $\alpha{xn}$ channels, corresponding to isotopes of W and Ta respectively, populated in the reaction of $^{16}$O+$^{165}$Ho are marked in the 
spectrum. The $pxn$ channel can be populated as an ER of compound nucleus and also from the decay product of Re (e.g. $^{177}$Re $\xrightarrow{\epsilon}$ 
$^{177}$W). The procedure to calculate the true weight of $pxn$ ERs from the cumulative cross-sections have been discussed by Cavinato \textit{et\ al.} 
\cite{cit21}. In the energy region of the present measurement, it has been found that majority of the $\gamma$-ray peaks corresponding to $pxn$ channels are
due to the decay of Re nuclei. The contribution of $pxn$ ERs, i.e. the direct decay product of compound nucleus falls within the error limit of respective 
cross-sections at different energies and thus have not been estimated rigorously. The $\alpha{xn}$ channels are populated either by incomplete fusion (ICF) 
or via the decay of compound nucleus(CF). To estimate the contribution of $\alpha{xn}$ channels arising from the CF residues, statistical model calculation 
has been performed using the PACE4 code \cite{cit22}.

\begin{figure}[t]
\centering
\includegraphics[clip, trim=0.0cm 0.1cm 1.0cm 1.6cm,width=0.475\textwidth, keepaspectratio]{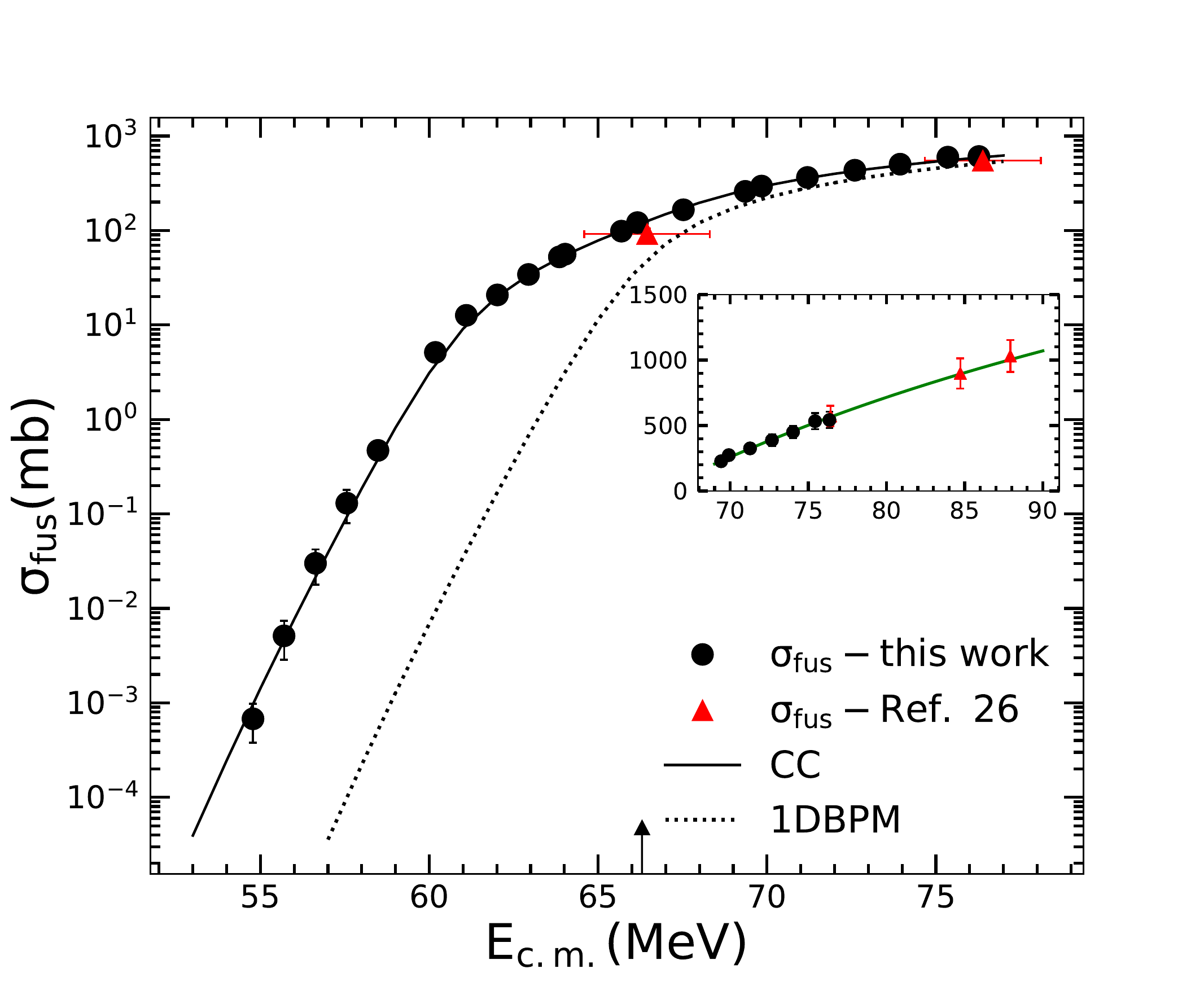}
\caption{Complete fusion excitation function measured in this experiment and the previous measurement has been compared with CC calculations using 
Woods-Saxon potential form. The solid lines are the results obtained from the CC calculations. The arrow corresponds to the Coulomb barrier for this system. (Inset) Fusion cross-section at above barrier energies are fitted with Wong formula to obtain the optical model potential parameters of the system. The solid line is the fitted result. }
\end{figure}

  \begin{table}[t]
\caption{Spectroscopic properties of the evaporation residues, resulting via CF process that have been used to calculate the CF cross-sections for the  $^{16}$O+$^{165}$Ho system.  } % title of Table
\centering % used for centering table
\begin{tabular*}{0.49\textwidth}{c @{\extracolsep{\fill}} ccccc} % centered columns (7 columns)
\hline\hline \\[-2.0ex]%inserts double horizontal lines
Residue & $T_{1/2}$(min) & $J^{\pi}$ & $E_{\gamma}$(keV) & $I^{\gamma}$(\%)\\[1.5ex]
%(MeV) &  (fm) & (fm) & (fm) & (fm) & (MeV) & (MeV) &  & (mb) \\[1.5ex] % inserts table
%heading
\hline % inserts single horizontal line
%23 & 0.8 & 0.8 & 1.09 & 1.1 & 0   \\ % inserting body of the table
$^{178}$Re(3$n$) & 13.2 & 3$^+$ & 237.0 & 44.5   \\
&  &  & 106.0 & 23.4 \\
$^{177}$Re(4$n$) & 14.0 & 5/2$^-$ & 196.9 & 8.4   \\
&  &  & 209.8 & 2.8 \\
$^{176}$Re(5$n$) & 5.3 & 3$^+$ & 240.3 & 54.0   \\
&  &  & 109.1 & 25.0 \\
$^{174}$Ta(${\alpha}3n$) & 68.4 & 3$^+$ & 206.5 & 60.0   \\
 \\ [-0.0ex] % [1ex] adds vertical space
\hline %inserts single line
\end{tabular*}
%\label{table:t1} % is used to refer this table in the text
\end{table}
To obtain a relatively better agreement between theoretical and experimental cross-sections, the neutron and  alpha potentials in the code have been modified according to Ref.\cite{cit23} and Ref.\cite{McFadden1966}, respectively. Fission barrier was fixed at 21.89 MeV following Ref.\cite{cit19} and the level density parameter "$k$" was taken to be 9. The other parameters in the code were set to the default values. Calculated cross section for each partial wave by coupled channel calculation (detailed discussion in next section)
has been fed as input to the PACE4 calculations. The production of $^{174}$Ta via ICF is energetically possible at higher energies (80-85 MeV) but at lower energies, it becomes unlikely that $^{174}$Ta will be populated as an ICF product. It has also been observed that within the energy range of the measurement, the experimental cross-sections of $\alpha{3n}$ channel ($^{174}$Ta)  reasonably agree with the statistical model, which exclusively deals with CF process. The experimentally measured 
cross-sections of other $\alpha{xn}$ channels are distinctly underestimated by the statistical model calculations. Similar trend has also been observed in 
Ref.\cite{cit19}. As the PACE4 code has been able to reproduce the cross-sections of $xn$ and $\alpha{3n}$ channels relatively well, it can
be inferred that all the other $\alpha{xn}$ channels, or at least an exceedingly significant part of them have been populated via incomplete fusion(ICF).  Fig.3 shows a comparison of the statistical model calculations with the measured ER channel cross-sections. 

The total fusion cross-sections have been determined by adding the measured cross sections of $xn$ and $\alpha{3n}$ channels and are shown in Fig. 4. 
Fusion cross section down to 678 nb has been measured in this work. Statistical errors, as well as errors ensuing from the measurement of beam current, 
target thickness and detector efficiency have been taken into account. 
\begin{figure}[t]
\centering
\includegraphics[clip, trim=0.5cm 0.5cm -1.0cm 0.4cm,width=0.46\textwidth, keepaspectratio]{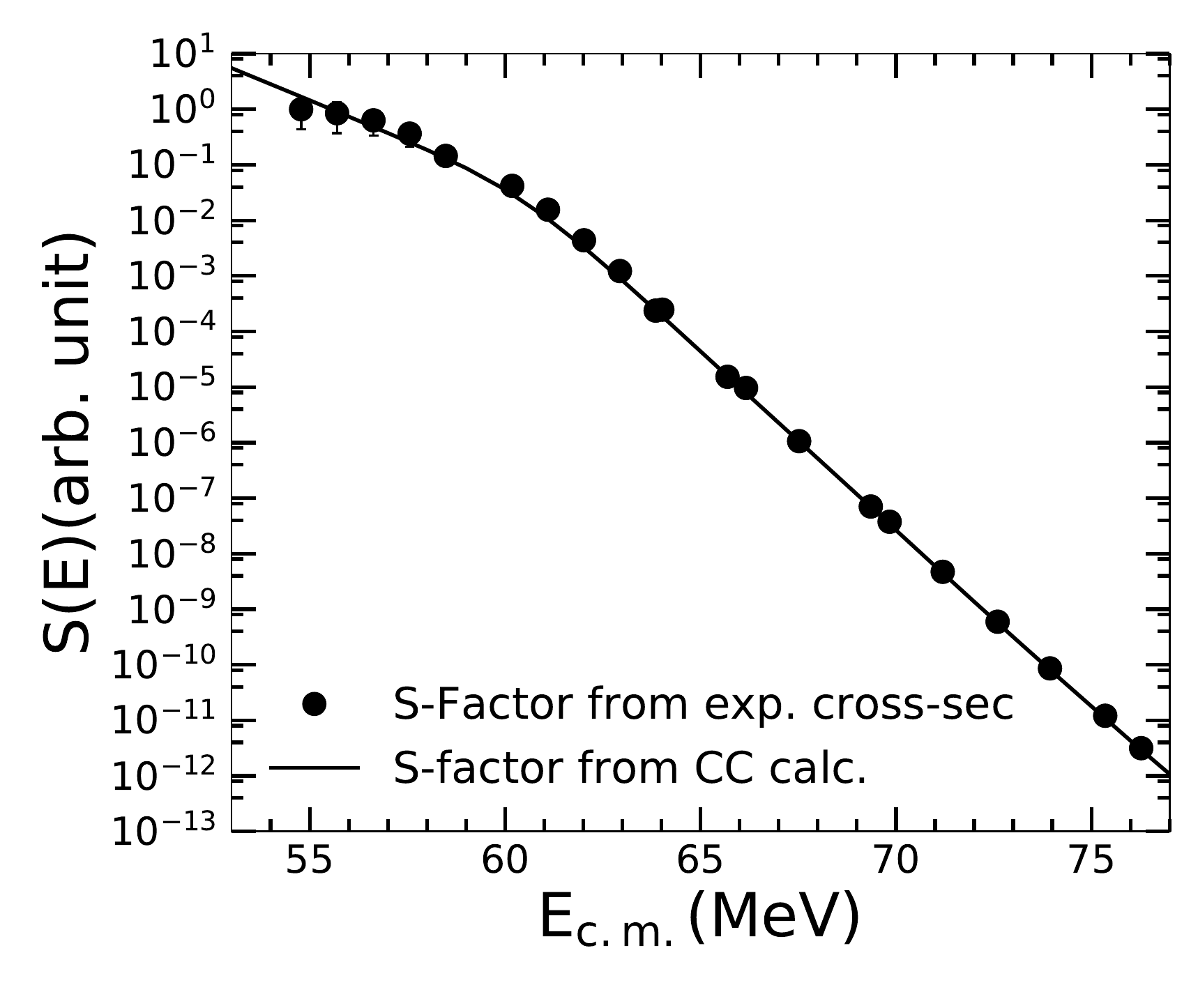}
\caption{Astrophysical S-factor for $^{16}$O+$^{165}$Ho system compared with S-factor obtained from CC calculation. }
\end{figure}
The astrophysical S-factor, often used in nuclear astrophysics to study the 
low-energy behavior of nuclear reactions, is defined as,
\begin{equation}
S(E)=E\sigma (E)exp(2\pi \eta),
\end{equation}
where E is the center-of-mass energy, $\sigma$ is the fusion cross-section and $\eta$=$Z_{1}Z_{2}e^{2}/\hbar v$ is Sommerfield parameter with $v$ being the beam
velocity. The experimental S-factor curve corresponding to the measured fusion 
cross-sections for the present system has been plotted in Fig. 5. Appearance of
a maximum in the S-factor curve has been presented in previous studies at an 
energy where the hindrance in fusion cross-section sets in. In the present 
study, although no clear maximum could be seen in the S-factor curve within the measured energy range, an indication of a change of slope at the lowest 
energy could be observed. 
An alternative representation, the logarithmic slope of the fusion excitation 
function is defined as \cite{cit1}, 
\begin{equation}
L(E)=d[ln(E \sigma)]/dE={\frac{1}{E\sigma}}\frac{d(E\sigma)}{dE}
\end{equation}
The values of $\mathrm{L(E)}$, extracted from the measured fusion cross
sections for $^{16}$O+$^{165}$Ho are plotted in Fig. 6. These representations 
are independent of any theoretical model and are alternative approaches to 
manifest any deviation in the slope of excitation function. From eqns. (2) and 
(3), one gets the relation,  
\begin{equation}
\frac{dS}{dE}=S(E)[L(E)-\frac{\pi{\eta}}{E}]
\end{equation}

The derivative $\frac{dS}{dE}$ becomes zero when S-factor becomes maximum, and 
from eqn. (3) one finds that this corresponds to the logarithmic derivative for
constant S-factor, $\mathrm{L_{cs}(E)}$, given by:
\begin{equation}
L_{cs}(E)=\frac{\pi{\eta}}{E}
\end{equation}
The dashed curve $\mathrm{L_{cs}(E)}$ for the present system is shown in Fig. 6 by the
dashed line. The experimental values of $\mathrm{L(E)}$ have been fitted with a
function $A+B/E^\frac{3}{2}$ \cite{cit25,cit26}, and is shown in Fig. 6 by the 
dotted line. The crossover point of the curves $\mathrm{L_{cs}(E)}$ and fitted 
$\mathrm{L(E)}$ corresponds to the maximum of the S-factor curve and is related
to the threshold energy ($\mathrm{E_{s}}$) for the occurance of fusion 
hindrance \cite{Back14}. In the measured energy regime of the present work, the
fitted $\mathrm{L(E)}$ curve does not intersect the $\mathrm{L_{cs}(E)}$ curve.
However, it can be seen from Fig. 6 that an extrapolation of the fitted 
$\mathrm{L(E)}$ curve intersects the curve $\mathrm{L_{cs}(E)}$ at the energy 
$\mathrm{E_{s}}=$52.78$\pm$0.78 MeV. The value of $\mathrm{E_{s}}$ calculated 
from the empirical equation \cite{cit27} is 53.17 MeV. The estimation of $\mathrm{E_{s}}$ 
obtained from the touching point configuration in adiabatic picture \cite{Ichikawa2009, Ichikawa2015} 
is 56.1 MeV.The present measurement 
has been performed down to E = 54.78 MeV, which is $\approx$ 2 MeV above 
the extrapolated $\mathrm{E_{s}}$ value and 1.3 MeV below the prediction from adiabatic model. 
This 
shows that the $\mathrm{E_{s}}$ value for $^{16}$O+$^{165}$Ho system will be lower than that 
predicted by the adiabatic model.

\section{COMPARISON WITH COUPLED CHANNELS CALCULATIONS}
\begin{figure}[t]
\centering
\includegraphics[clip, trim=0.2cm 0.5cm 0.0cm 0.4cm,width=0.48\textwidth, keepaspectratio]{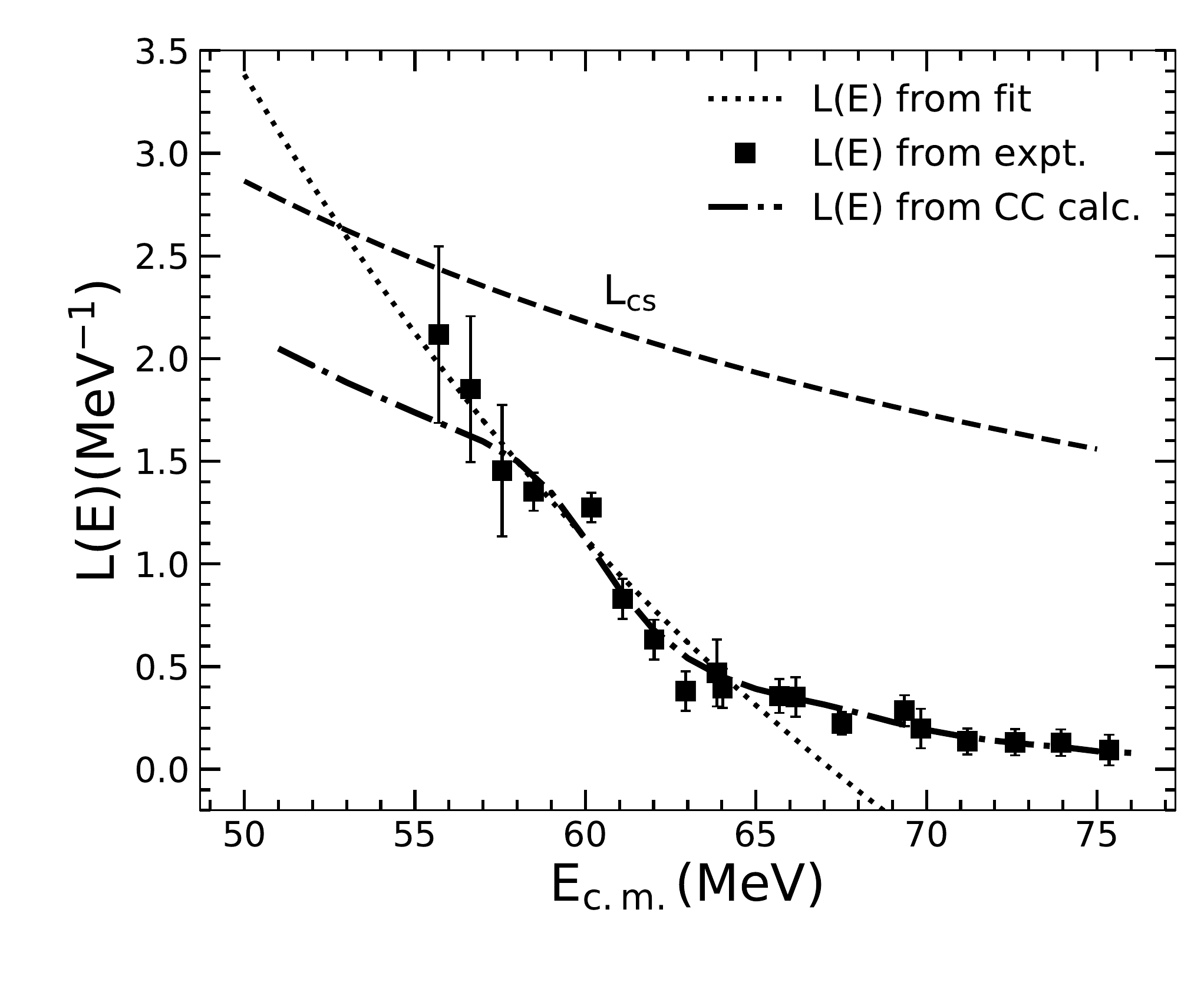}
\caption{Logarthmic derivative L(E) of the experimental fusion excitation function compared with CC 
calculations(dashed dot lines). The extrapolation of 
L(E) (dotted lines) intersect the $\mathrm{L_{cs}}$ curve at 52.78 MeV. }
\end{figure}
The data obtained in the present work have been analyzed in the framework of 
coupled channels (CC) calculations, using the code CCFULL \cite{cit28}. These
calculations require an initial set of potential parameters. They were obtained
by fitting the fusion cross sections well above the barrier using Wong's 
formula\cite{cit30}, as the fusion cross sections in this energy regime are 
expected to be fairly insensitive to the form or magnitude of
the couplings. The nuclear potential was taken to be of Woods-Saxon form,
\begin{equation}
V_{n}(r)=\frac{-V_0}{1+exp[(r-r_{0}A_{P}^{1/3}-r_{0}A_{T}^{1/3})/a]}
\end{equation}
where $V_{0}$ is the depth, $r_{0}$ is the radius parameter, and $a$ is the 
diffuseness of the nuclear potential. 
The potential parameters were obtained by fixing $a$ to be 0.63 fm, and varying
$r_{0}$ and $V_{0}$ to obtain a good fit to the high energy part of
the cross sections. The parameters thus obtained are: $V_0=$102 MeV, $r_0=$1.15
fm and $a_0=$0.63 fm. The inset in Fig. 4 shows the resulting fit. The potential parameters for the 
same system has also been compared  with Ref. \cite{Hagino2019}; both results  are found to be identical.

The CCFULL calculations in the no-coupling limit are shown by the dotted curve 
in Fig. 4, and are seen to underpredict the data, suggesting strong effects
of deformation in the target nucleus. The CCFULL code estimates the effects of 
deformation by linear as well as non linear coupling to the pure rotational 
bands of the deformed nucleus. The target nucleus $^{165}$Ho is deformed with a 
valence proton. The rotational states of $^{165}$Ho may be considered to be
built up by the coupling of the unpaired valence proton particle (or proton 
hole) with the 0$^+$,2$^+$,4$^+$,...rotational states of the neighboring 
even-even nucleus $^{164}$Dy (or $^{166}$Er). To remain within the model 
space of CCFULL, the excitation energies and deformation parameters for the 
target nuclues $^{165}$Ho were taken to be the averages of those of the 
neighboring even-even nuclei $^{164}$Dy and $^{166}$Er \cite{cit32,cit33}. The 
resulting ground state rotational band upto 12$^+$ state ($\beta_2$=0.32
, $\beta_4$=0.02 ) was included in the calculations. The 
results of these 
calculations are shown by the solid curve in the figure. They are seen to be in
fairly good agreement with the data, except the lowest energy data which 
appears to be slightly below the calculated value. 

The solid curve in Fig. 5 shows the calculated astrophysical S-factor values 
extracted from the CC cross sections and are found to agree well with the
experimental value, except the lowest energy point. Although no clear maximum 
in the experimental S-factor plot or deviation from the the CC calculations 
could be observed in the measured energy range, an indication of a small 
deviation could be seen at the lowest energy measured. The dot-dashed curve in
Fig. 6 shows the calculated $\mathrm{L(E)}$ values extracted from the CC cross 
sections and and are also seen to agree well with the experimental values, 
although the trend of the two extreme low energy points show an indication of 
possible deviation at still lower energies.

\begin{figure}[t]
	\centering
    \includegraphics[clip, trim=1.25cm 2.0cm -0.3cm 0.6cm,width=0.455\textwidth, keepaspectratio]{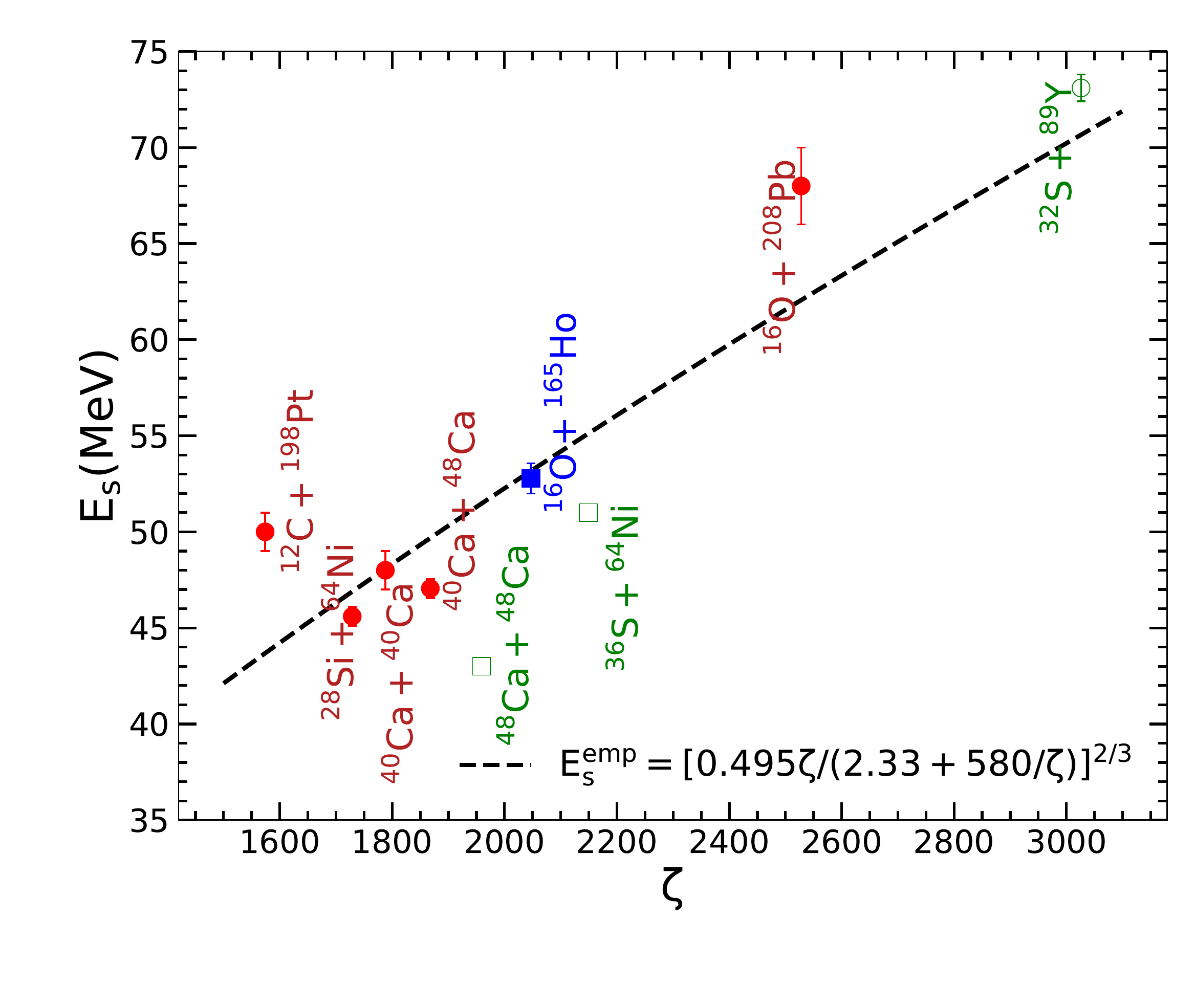}
    
 	\caption{Systematic representation of $\mathrm{E_s}$ as a function of system parameter 
$\mathrm{\zeta}$. The solid circles are measured and hollow circles are extrapolated values of 
$\mathrm{E_s}$, taken from the tabulated values of Ref.\cite{cit13,Back14,cit25}. The solid 
square represents the extrapolated value in this work. The hollow squares correspond to 
Ref.\cite{cit7,cit35} where the value of $\mathrm{E_s}$ can not be obtained either by extrapolation 
or direct measurements and an upper bound has been mentioned in Ref.\cite{Back14}.}
\end{figure}
\section{DISCUSSION}
For a systematic study of the energy onset of fusion hindrance, the value of 
$\mathrm{E_{s}}$ predicted for the system $^{16}$O+$^{165}$Ho in the present 
work has been compared in Fig. 7 with the $\mathrm{E_{s}}$ values of other 
heavy systems as a function of the system parameter
$\mathrm{\zeta=Z_pZ_t\sqrt{\frac{A_pA_t}{A_p+A_t}}}$. The parameters 
$\mathrm{A_p,Z_p,A_t,Z_t}$ are the mass number and atomic number of projectile 
and target. In this comparison, we have considered only the systems in the 
$\mathrm{\zeta}$ range $\sim$ 1500-3000, for which $\mathrm{E_{s}}$ are available 
in the literature. The parameter $\mathrm{\zeta}$ contains information 
of the mass and Coulomb barrier of the target-projectile system. The 
dashed line in the figure corresponds to the empirical form of 
$\mathrm{E_s}$ \cite{cit27}. The extrapolated value of $\mathrm{E_{s}}$ obtained
in the present work lies very close to the result obtained from the empirical 
form $\mathrm{E_{s}^{emp}=[0.495\zeta/(2.33+580/\zeta)]^{2/3}}$; although it should be pointed out 
that an actual measurement down to the 
predicted threshold energy will confirm the result. On the other hand, the $\mathrm{E_{s}}$ limit, 
obtained from the adiabatic model, has been crossed but hindrance for this system was not experimentally 
observed within the measured energy range.

\begin{figure}[h]
	\centering
    \includegraphics[clip, trim=1.35cm 0.9cm 0.0cm 0.0cm,width=0.47\textwidth, keepaspectratio]{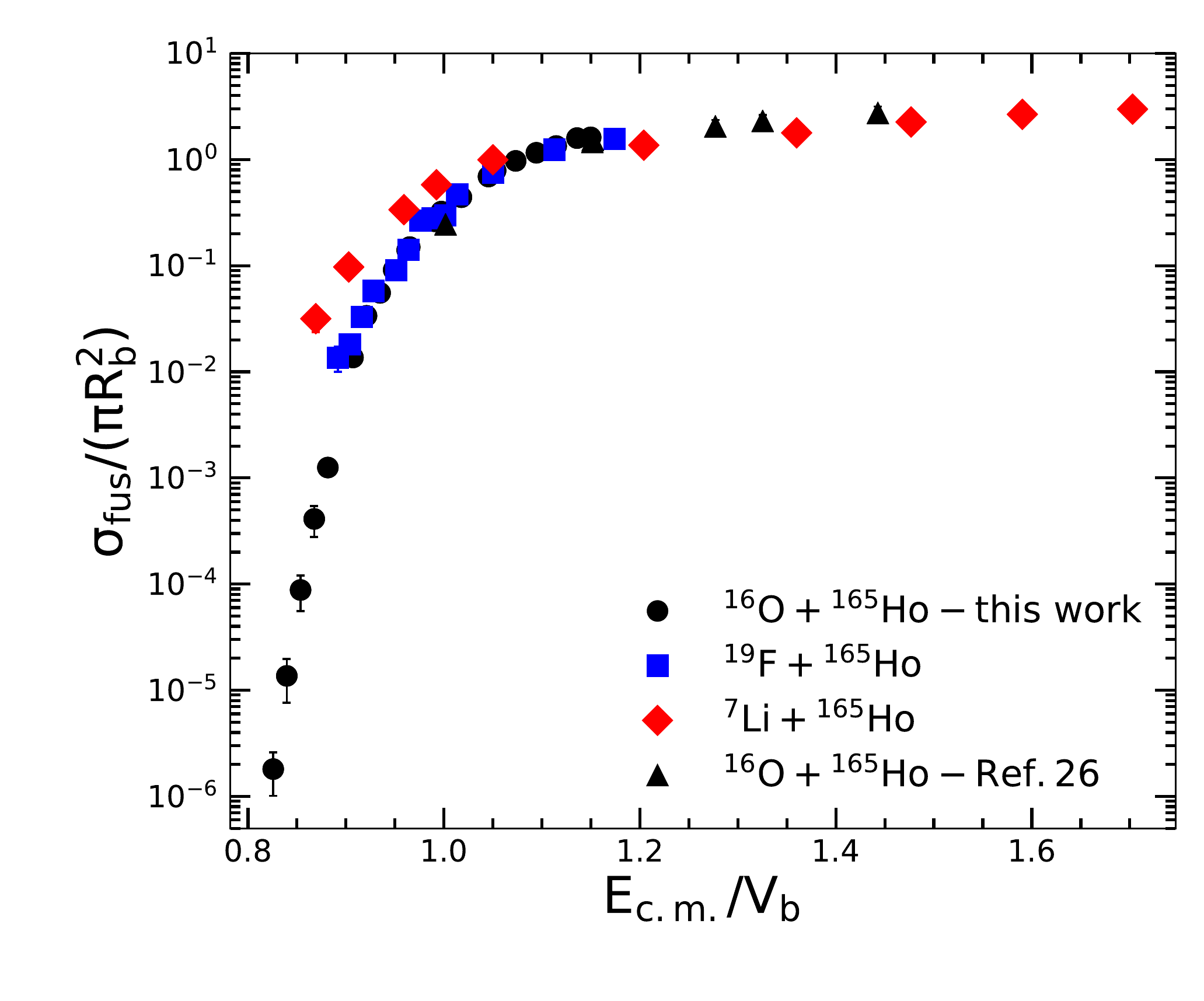}
    
  	\caption{Comparison of normalized fusion excitation function between $^{16}$O+$^{165}$Ho, $^{19}$F+$^{165}$Ho \cite{cit34}, and $^{7}$Li+$^{165}$Ho \cite{Tripathi2002} system.}
\end{figure}
A systematic comparison of fusion excitation functions have been carried out for
systems involving the target nucleus $^{165}$Ho and different projectiles and is
shown in Fig. 8. The systems involving $^{165}$Ho target nucleus, for which sub-barrier
fusion cross sections have been reported in literature are $^{19}$F+$^{165}$Ho  \cite{cit34} and 
$^{7}$Li+$^{165}$Ho \cite{Tripathi2002}. The projectile $^{16}$O in the present system is a stiff 
nucleus, while $^{19}$F is relatively heavier and less bound than $^{16}$O, and $^{7}$Li is a 
well known weakly bound stable nucleus.  For comparison of different projectile-target systems, 
the fusion excitation functions have been plotted in a reduced scale.  Barrier radii $\mathrm{R_b}$=$\mathrm{r_0(A_p^\frac{1}{3}+A_t^\frac{1}{3})}$ and Coulomb barrier $\mathrm{V_b}$ for each system was obtained by performing 1DBPM model calculation, using the Aky{\"u}z-Winther parametrization of Woods-Saxon potential \cite{cit29}.

\begin{table}[h]
\caption{$V_b$ and $R_b$ of the systems that have been used in the reduced plot (Fig. 8).  } % title of Table
\centering % used for centering table
\begin{tabular*}{0.48\textwidth}{c @{\extracolsep{\fill}} cccccc} % centered columns (7 columns)
\hline\hline \\[-2.0ex]%inserts double horizontal lines
& System & & $V_b$ (MeV) & & $R_b$ (fm) \\[1.0ex]
%(MeV) &  (fm) & (fm) & (fm) & (fm) & (MeV) & (MeV) &  & (mb) \\[1.5ex] % inserts table
%heading
\hline % inserts single horizontal line
%23 & 0.8 & 0.8 & 1.09 & 1.1 & 0   \\ % inserting body of the table
&$^{16}$O+$^{165}$Ho & & 66.34 & & 10.93   \\
&$^{19}$F+$^{165}$Ho & &72.24 & & 11.26    \\
&$^7$Li+$^{165}$Ho & &25.0 & & 10.76    \\
 \\ [-2ex] % [1ex] adds vertical space
\hline %inserts single line
\end{tabular*}
%\label{table:t1} % is used to refer this table in the text
\end{table}
Fig. 8 shows that the reduced fusion excitation functions for the three systems 
overlap reasonably well with each other at above barrier energies. But there are
no reported fusion cross sections at deep-sub-barrier energies for the systems 
$^{19}$F+$^{165}$Ho and $^{7}$Li+$^{165}$Ho. At sub-barrier energies, the fusion cross-sections for 
$^{19}$F+$^{165}$Ho and $^{16}$O+$^{165}$Ho agree fairly well with each other; while the fusion cross-sections 
for $^{7}$Li+$^{165}$Ho are enhanced in the reduced scale. It would be interesting to see
how the deep sub-barrier fusion cross sections for the two systems compare with
the present $^{16}$O+$^{165}$Ho system. 

\section{SUMMARY}
The fusion cross-sections for $^{16}$O+$^{165}$Ho have been measured down to 
678 nb, from above-barrier to deep sub-barrier energies. The present 
measurement agrees with the earlier reported data \cite{cit19}, in the 
overlapping above-barrier energy region. Statistical model calculation of 
different evaporation residues 
occurring from CF yielded similar results with experimental data. The 
experimental CF cross-sections have been well reproduced by the coupled 
channels calculations. No clear evidence of fusion hindrance has been 
observed in the fusion excitation function in the measured energy range of the 
present work. However, the trend of the extreme low energy points in the 
S-factor and $\mathrm{L(E)}$ plots show an indication of possible deviation at 
still lower energies. The extrapolation of $\mathrm{L(E)}$ suggests the onset 
of fusion hindrance at 52.78$\pm$0.78 MeV, approximately 2 MeV lower than the
lowest energy measured in the present work, in accordance with the prediction 
from systematics. Further measurement at lower energies may confirm the result.

\begin{acknowledgments}
We sincerely thank the staff of BARC-TIFR Pelletron facility for an 
uninterrupted supply of beam. We 
would like to extend our gratitude towards Dr. Sanjoy Pal of TIFR, Mumbai; for 
helping with the digital data acquisition. 
\end{acknowledgments}

\end{document}